\documentclass{egpubl}
\usepackage{ICAT-EGVE2021}

 \WsPaperJoint         

\usepackage[T1]{fontenc}
\usepackage{dfadobe}  

\usepackage{cite}  
\BibtexOrBiblatex
\electronicVersion
\PrintedOrElectronic
\ifpdf \usepackage[pdftex]{graphicx} \pdfcompresslevel=9
\else \usepackage[dvips]{graphicx} \fi

\usepackage{egweblnk} 

\usepackage{epsfig} 
\usepackage{mathptmx} 
\usepackage{times} 
\usepackage{amsmath} 
\usepackage{amssymb}  
\usepackage{comment}
\usepackage{microtype}
\usepackage{xcolor}

\title[Electrotactile Feedback For Enhancing Contact Information in Virtual Reality]%
      {Electrotactile Feedback For Enhancing \\ Contact Information in Virtual Reality}
      
\author[S. Vizcay, P. Kourtesis, F. Argelaguet, C. Pacchierotti \& M. Marchal]
{\parbox{\textwidth}{\centering S. Vizcay$^{1}$\orcid{0000-0002-1837-5607}, P. Kourtesis$^{1}$\orcid{0000-0002-2914-1064}, F. Argelaguet$^{1}$\orcid{0000-0002-6160-8015}, C. Pacchierotti$^{2}$\orcid{0000-0002-8006-9168} and M. Marchal$^{3}$\orcid{0000-0002-6080-7178}}
\\
{\parbox{\textwidth}{\centering $^1$Inria, Univ Rennes, IRISA, CNRS – 35042 Rennes, France.\\
$^2$CNRS, Univ Rennes, IRISA, Inria – 35042 Rennes, France.\\
$^3$Univ Rennes, INSA, IRISA, Inria, CNRS – 35042 Rennes, France and Institut Universitaire de France.}}
}

\begin{document}


\maketitle
\begin{abstract}
This paper presents a wearable electrotactile feedback system to enhance contact information for mid-air interactions with virtual objects. In particular, we propose the use of electrotactile feedback to render the interpenetration distance between the user's finger and the virtual content is touched. Our approach consists of modulating the perceived intensity (frequency and pulse width modulation) of the electrotactile stimuli according to the registered interpenetration distance. In a user study (N=21), we assessed the performance of four different interpenetration feedback approaches: electrotactile-only, visual-only, electrotactile and visual, and no interpenetration feedback. First, the results showed that contact precision and accuracy were significantly improved when using interpenetration feedback. Second, and more interestingly, there were no significant differences between visual and electrotactile feedback when the calibration was optimized and the user was familiarized with electrotactile feedback. Taken together, these results suggest that electrotactile feedback could be an efficient replacement of visual feedback for enhancing contact information in virtual reality avoiding the need of active visual focus and the rendering of additional visual artefacts.


\end{abstract}  


\section{Introduction}

Immersive virtual reality (VR) is becoming a common platform for education and training, where precision and accuracy are important aspects \cite{carruth2017virtual,jensen2018review}. Visual feedback is important for contact accuracy and precision \cite{Triantafyllidis2020}.  However, when interacting in VR, if no other measures are adopted, the user's avatar might interpenetrate the virtual content, which results in significantly reducing the accuracy, precision, and the realism of the interaction
To avoid such an undesired behavior and facilitate mid-air interactions in VR  a popular technique is the \textit{God-object Method}, where the virtual hand is constrained by the collision with the virtual object(s) \cite{ZillesGOD,SagardiaGOD}. Using the \textit{God-object Method}, there is a decoupling of the visual representation of the user's avatar (e.g., the virtual hand) from its actual position by, e.g., constraining the avatar on the virtual object's surface even when the actual tracked position would be inside the object~\cite{Prachyabrued2016}.%

However, using this popular technique, while the virtual hand is rendered on the collided object's surface, the user's physical hand is still able to move in the real environment unconstrained (i.e., penetrating the virtual object's coordinates), which results in failing to perceive the inaccuracy and imprecision of the contact \cite{ZillesGOD,SagardiaGOD}. Considering the dominance of visual information, a visual interpenetration feedback could provide the required information to facilitate contact precision and accuracy. Nevertheless, providing additional visual cues, would occlude other virtual surfaces and render visual artefacts in the virtual environment. Visual artefacts negatively affect contact precision and accuracy \cite{qian2015virtual}. To cope with this limitation, a haptic feedback proportional to the interpenetration may provide the user with information regarding the contact precision and accuracy. This haptic information hence indicates the user's interpenetration in the virtual content~\cite{meli2018combining}, which would enable to somehow mimic the behavior of real contacts, where interpenetrating deeper in a virtual object elicits stronger tactile sensations.%

Haptic VR interfaces provide vibrotactile, pressure, or skin stretch feedback sensations \cite{Pacchierotti2017}. However, these displays need to embed one or more mechanical actuators that move an end-effector in contact with the user's skin, limiting their wearability, portability, and comfort~\cite{Pacchierotti2017}.
In this respect, electrotactile haptic stimulation has been recently recognized as a promising technique to provide rich tactile sensations across the hand or arm through a single wearable actuator \cite{CHOUVARDAS}.
Applying electrical stimulation to the skin elicits tactile sensations by directly activating the sensory nerve endings, de facto bypassing the skin mechanoreceptors. 
The electrical current is produced by a local polarization near
the skin, generated by two or more electrotacile
interfaces (electrodes).  
With respect to other tactile solutions, electrotactile stimulation only requires very thin electrodes placed at the sensation point (e.g., the fingertip), providing a wearable and comfortable solution to tactile haptics.%

In a VR pilot study (N=3), electrotactile feedback provided haptic information pertinent to heat, texture, jolt, and texture \cite{pamungkas2016electro}. In a small size user study (N=10), electrotactile feedback was able to provide the VR user with tactile sensations such as  roughness, friction, and hardness \cite{7892236}. In a VR user study (N=19), electrotactile feedback was successful in improving grasping \cite{hummel2016lightweight}. Despite its promising features, the benefits of electrotactile feedback in VR interaction are still under-investigated.
This paper presents the design and evaluation of a wearable electrotactile feedback system to enhance contact information for mid-air interactions with virtual objects.

The main contributions of our work can be summarized as follows:
\begin{itemize}
    \item Design of an electrotactile feedback technique for enhancing contact information in VR. A technique where the intensity of the electrotactile feedback is proportional to the interpenetration (i.e., distance between real and virtual hand).%
    \item A user study (N=21) for evaluating electrotactile interpenetration feedback's performance against no interpenetration feedback, a visual interpenetration feedback, and a visual-electrotactile interpenetration feedback.%
\end{itemize}

\section{Related work}
\emph{Wearable haptics in VR.}
Popular techniques to provide rich wearable haptic feedback are through moving platforms, that can orient and/or translate on the skin, pin-arrays, shearing belts and tactors, pneumatic jets, and balloon-based systems~\cite{Pacchierotti2017}.
Frisoli et al.~\cite{frisoli2008fingertip} were among the first to present a fingertip haptic display for improving curvature discrimination using a moving platform. It was composed of a parallel platform and a serial wrist.
Gabardi et al.~\cite{Gabardi2016} improved this device and applied it in VR interaction. The pose of the user's finger was tracked through external optical sensors and collisions between the finger avatar and the virtual surface were rendered by the device.
Minamizawa et al.~\cite{minamizawa2007wearable} presented a wearable fingertip device for weight rendering in VR. It consisted of two DC motors moving a belt in contact with the user's fingertip. A similar device was also used in~\cite{PaSaHuMePr-hs16} for multi-finger manipulation of objects in VR and in~\cite{salazar2020altering,de2018enhancing} for augmenting tangible objects in VR.
More recently,
Girard et al.~\cite{gosselin2017design} developed a wearable fingertip device capable of rendering 2-DoF skin stretch stimuli in VR.
The considered use cases comprised tapping on a virtual bottle and feeling the texture and weight of a virtual object.

\emph{Electrotactile haptics.}
Due to the difficulty in designing effective interaction paradigms, electrotactile interfaces have rarely been  used in VR scenarios.
Jorgovanovic et al.~\cite{jorgovanovic2014virtual} evaluated the capabilities of human subjects to control grasping force in closed loop using electrotactile feedback. It was tested in an experiment of virtual robotic grasping. 
More recently, a pilot study by Pamungkas and Ward~\cite{pamungkas2016electro} used electrotactile stimulation on the hand to convey the sensation of contacts in VR. Use cases comprised playing with a virtual bouncing ball, setting up a camp fire, feeling different surface textures, and shooting a gun. Hummel et al.~\cite{hummel2016lightweight} developed an electrotactile delivery system composed of a small tactor with eight electrodes for each finger.
They tested it in three VR assembly tasks for remote satellite servicing missions, which included pressing a button, switching a lever switch, and pulling a module from its slot.
Finally, Yem et al.~\cite{Yem2018} employed electrotactile stimulation to elicit flexion/extension illusory movements to the user's fingertip. Doing so, the system was able to induce haptic feedback sensations, providing information about virtual objects' stiffness and stickiness properties.

\emph{Contact information in VR.}
In the last years, researchers focused on the sensation of making and breaking contact~\cite{tinguy2020weatavix,chinello2017three,chinello2019modular,young2020compensating}, as it is proven to be an important factor for immersion in VR.
Kuchenbecker et al.~\cite{kuchenbecker2008touch} endowed a desktop kinesthetic interface with a passive tactile end-effector. The kinesthetic feedback provided by the desktop device bends the internal springs of the tactile end-effector and brings a shell in contact with the user's finger, providing the sensation of making/breaking contact with the virtual object.
In a similar fashion, Gonzalez et al.~\cite{gonzalez2015smooth} attached a ring of optical sensor to the end-effector, being able to measure the distance of the fingertip from the ring and adjust the transition between free space and contact
Finally, Prachyabrued and Borst~\cite{Prachyabrued2016} evaluated the impact of visual feedback in pinch grasping a virtual object, where best performance was obtained when showing both hand models (tracked and constrained) and when coloring the fingers of the object being grasped.

To the best of our knowledge, the effects of using electrotactile feedback for enhancing contact information in VR mid-air interactions has scarcely been studied.

\section{Electrotactile Interpenetration Feedback}
\label{sec:contactrendering}
\label{sec:feedbackdesign}
The tactile sensation elicited by electrotactile stimulation can be controlled by both, the parameters chosen for the design of the electrodes and the parameters of the delivered electrical signal~\cite{Kaczmarek2017,folgheraiter2008multi,kajimoto1999tactile,lozano2009electrotactile}.
Individual pads in an electrode can be disabled or enabled (enabled as either cathode or anode) meanwhile signal's intensity, pulse width and frequency can be increased/decreased to apply some modulation.
Pad selection determines the location where the stimulus is going to be applied.

Changes in the current intensity is the most common way to increase/decrease the strength of the sensation, but the feeling is highly variable across subjects and relatively small increases can quickly elicit uncomfortable sensations. Changes in the pulse width and frequency can also elicit changes in the perceived strength of the sensation, although less significantly than when altering the current intensity~\cite{Aiello1998}.
Changes in the frequency are also employed to increase/decrease 
how fast the tickles are delivered.

Given the disadvantages of intensity modulation, we fixed the intensity value for each participant, chosen after a brief calibration so as to avoid any undesired uncomfortable/painful sensation. We chose pulse width modulation to control the strength of the sensation and frequency modulation to control its speed.
We chose to place the electrodes at the fingertip, as it is the area of the body most often used for pointing, grasping, and manipulating objects ~\cite{Pacchierotti2017}.
To produce the stimulation at the center of the finger pad, we select an anode-cathode configuration with the widest possible dynamic range for the given electrode.
As mentioned before, we modulate the electrotactile interpenetration feedback proportionally to the distance between the constrained user's fingertip avatar (i.e., the object surface) and its actual position as tracked by the system. In other words, both the perceived intensity and speed of the electrotactile sensation increase as the user's physical hand interpenetrates the virtual object.%

\begin{figure}[htbp]
 \includegraphics[width=\columnwidth]{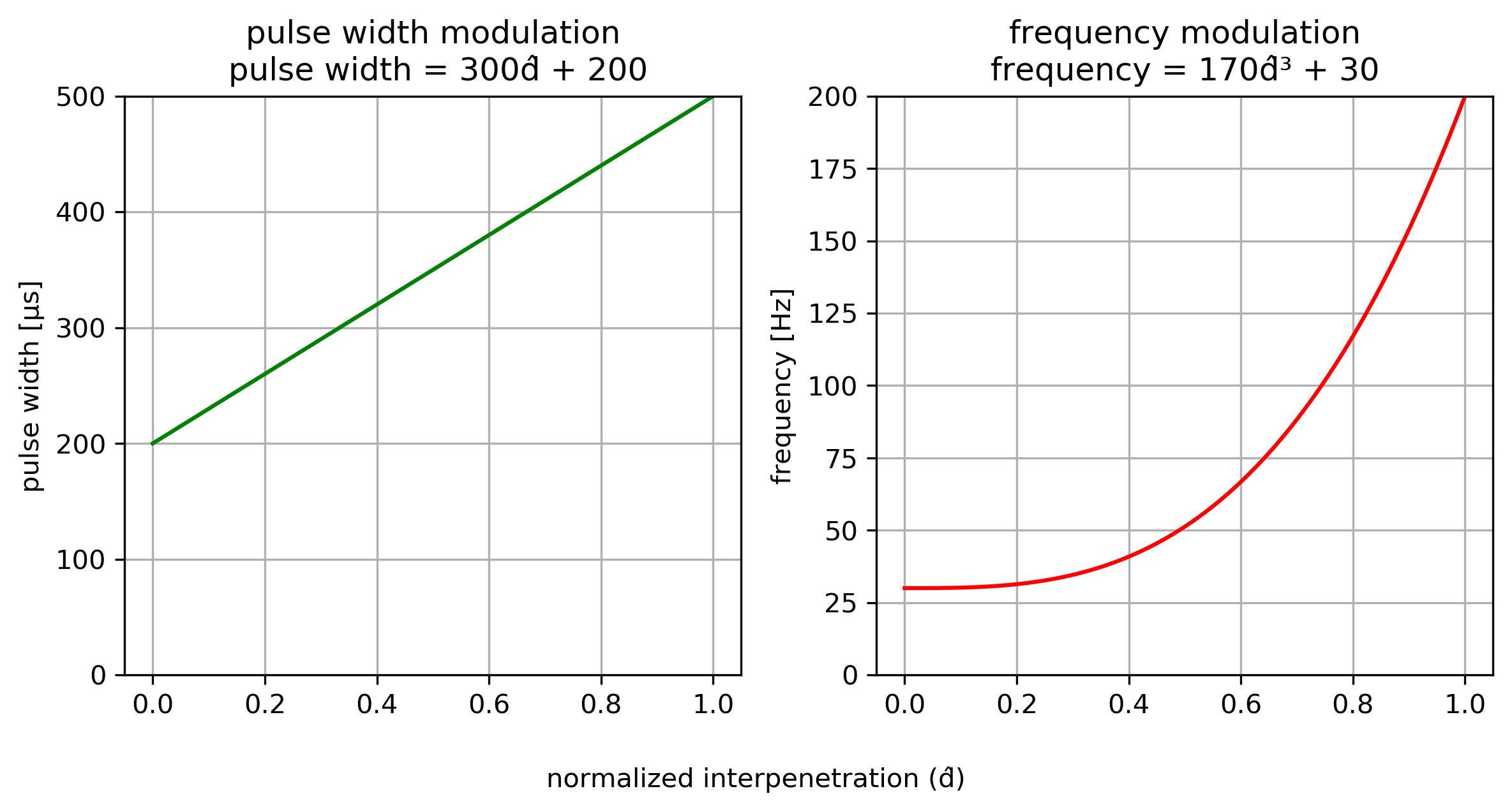}
 \vspace{-6mm}
 \caption{electrotactile interpenetration feedback modulation via pulse width (left) and frequency (right).}
 \label{fig:electrotactile-modulation}
\end{figure}

To achieve this effect, the pulse width is modulated linearly from 200~$\mu$s to 500~$\mu$s. On the other hand, we considered the perception of frequency to be logarithmic~\cite{Varshney2013} and
we apply a gamma correction so as to linearize how fast the tingles are perceived.
The proposed frequency modulation allows us to explore a range of frequencies from 30~Hz to 200~Hz, which is the maximum frequency of the device. Fig.~\ref{fig:electrotactile-modulation} shows the two inputs with respect to the normalized interpenetration (1: maximum, 0: zero, see \autoref{sec:system} for the details). Choices regarding the range of stimulation and the active pads have been done after a short preliminary study, so as to provide the most comfortable tactile sensations to the finger pad.%
\subsection{Electrical Stimulator and Electrodes}
\label{sec:stimDetails}
Our custom electrotactile stimulator has 32 channels, to which up to 5 electrodes can be connected simultaneously. Each channel may be defined either as a cathode or as an anode.
The stimulator produces a biphasic cathodic stimulation through square waves.
Electrical intensity and pulse width can be set per each channel in the ranges of [0.1, 9]~mA and [30, 500]~$\mu$s, respectively. The stimulus frequency can be set in the range [1, 200]~Hz.

\begin{figure}[t]
 \includegraphics[width= \columnwidth]{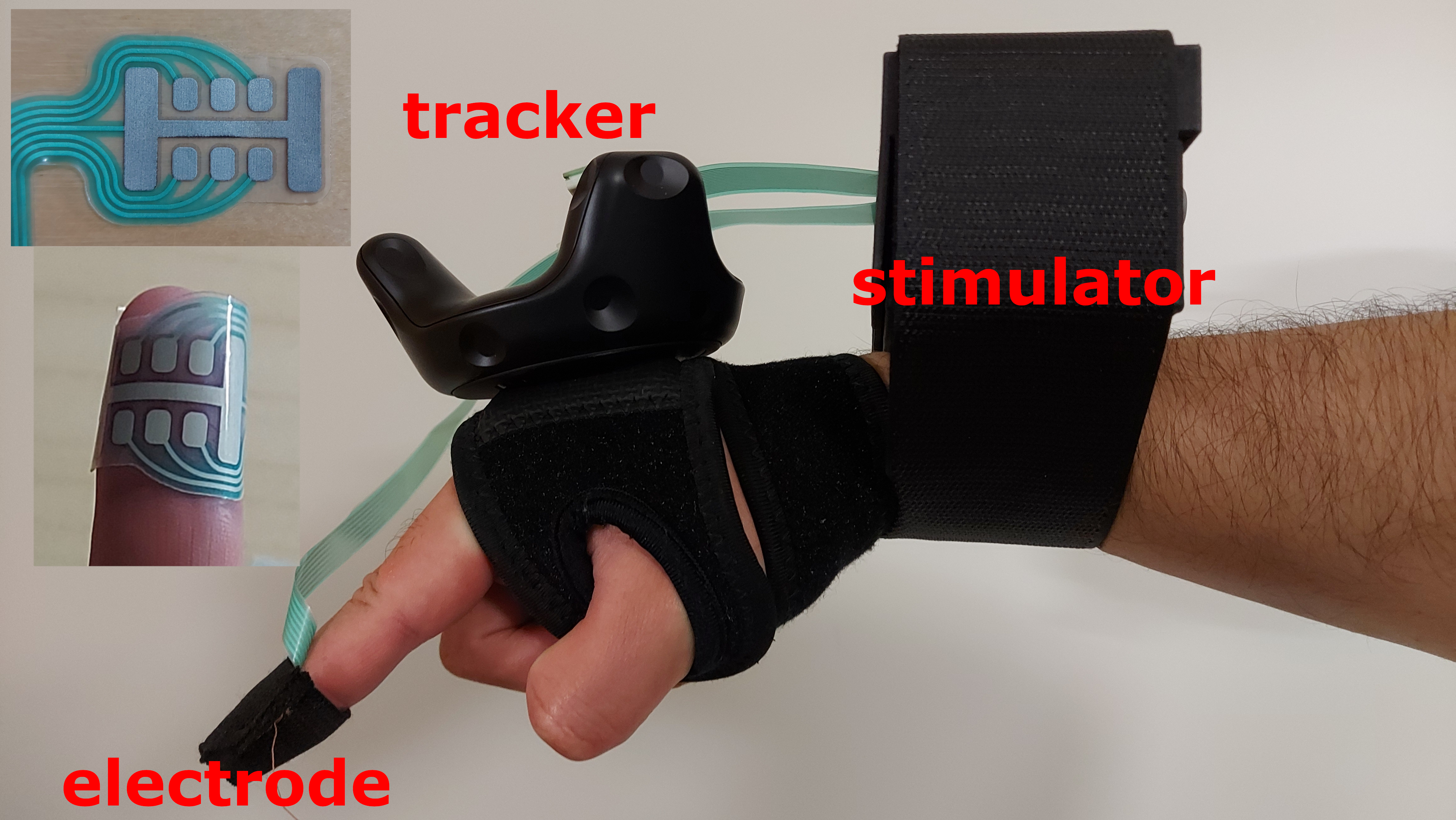} 
 \caption{Experimental apparatus. The electrical stimulator is attached to the forearm and the electrode is placed in contact with the finger pad. The two insets represent detailed views of the electrode. The user's hand is tracked by a HTC Vive Tracker.}
 \label{fig:system}
\end{figure}

We used electrodes whose cathodes are
distributed in a matrix array of 2 rows by 3 columns, surrounded by two larger lateral anodes that are interconnected by a segment going through both rows of cathodes (see the insets in the top-left corner of~\autoref{fig:system}). This particular layout has being designed to widen the dynamic range of sensations.%
\subsection{System Design}
\label{sec:system}
The interpenetration of the finger avatar into the virtual content drives our pulse width and frequency modulation. After a series of pilot user studies, we determined that, for a cube of edge of 15~cm, 3~cm represents the maximum interpenetration achieved in most interactions. For this reason, we normalized the interpenetration $d$ by this value, i.e., a normalized interpenetration $\hat{d}$ of 1 means a true interpenetration into the virtual object of 3~cm.%

The position of the participant's hand is tracked using a HTC Vive Tracker attached to the back of the hand. The \emph{physics-based} hand avatar (physics-based as its interactions are handled by the physics engine) collides with the other objects in the virtual environment while trying to follow the position indicated by the tracker. As described at the beginning of \autoref{sec:contactrendering}, this avatar resolves all virtual collisions and avoids any undesired virtual interpenetration. Whenever it comes into contact with a virtual object, its motion is constrained to stay on the object's surface, regardless of how much the user's hand moves into the object. As detailed above, this discrepancy between the user's real hand and its avatar drives our electrotactile interpenetration feedback.%
\section{Evaluation}
This section describes a user study evaluating the proposed electrotactile interpenetration feedback system. 
Our main objective was to assess the potential advantages and disadvantages of using electrotactile interpenetration feedback while performing a task requiring a high level of accuracy and precision. 
%
In particular, we focused on a task in which the user is asked to contact a virtual surface with his/her index finger while minimizing the amount of interpenetration between the virtual surface and the real index finger.
In such a task, when the virtual hand avatar cannot interpenetrate virtual surfaces, it is difficult for the user to precisely assess the amount of interpenetration between his/her finger and the virtual surface without any additional feedback.

Our main hypothesis is that additional interpenetration feedback, either visual and/or electroactile, would increase the awareness of the user, effectively decreasing the amount of interpenetration. Yet, electrotactile interpenetration feedback will provide a more coherent experience.
In addition to consider the electrotactile interpenetration feedback method presented in the previous section, for comparison purposes, we also modulated the availability of visual cues.
%
First, we considered a visual interpenetration feedback technique. Second, we considered the interaction with two different virtual objects, a fully-opaque virtual cube and a wireframe-only cube. We also considered the latter so as to limit the visual cues which could anticipate the contact with the virtual object.%

Hence, the independent variable (IV) $Interpenetration$ $Feedback$ has four levels: 1) No Interpenetration Feedback; 2) Visual Interpenetration Feedback; 3) Electrotactile Interpenetration Feedback, 4) Visuo-electrotactile Interpenetration Feedback (i.e., interpenetration information is provided both visually and haptically). The four levels of the IV are examined in two object rendering methods: 1) fully-opaque and 2) wireframe-only. In the following, we detail how these techniques are implemented in our experimental study.%


\subsection{Experimental Task}
The task required participants to touch a virtual surface (a cube of size $15\times15\times15$~cm) which was located on a virtual table (see \autoref{fig:environment}).
Participants started the experience in a seated position in front of said virtual table. The cube was placed on the user's left side, while a resting area was located on the user's right side.
For each trial, participants were asked to start with their right hand in the resting area, to move to touch with their index finger the top side of the virtual cube precisely while keeping a steady contact for three seconds, and then they were notified that should return their hand again to the resting area.
%
Participants were instructed to minimize the interpenetration between the virtual surface and their real index finger while ensuring a continuous contact with the cube.
To ensure that all participants kept a consistent timing, auditory feedback (a short beep sound) was used to notify the users when they had to start the trial and when the three seconds of contact were achieved.
During the whole interaction, participants were asked to keep their hand in a closed fist, with the index finger fully extended for facilitating a similar posture and an approximate collocation between the physical and the virtual hand. The experimenter was prompting each participant to have this posture when a new block was starting, while during the breaks between the blocks, the participant could have a rest and move her/his hand and fingers freely or just relax, which assisted with avoiding a postural fatigue.%
\subsection{Electrotactile Interpenetration Feedback Calibration}
\label{sec:calibration}
The calibration of the electrotactile interpenetration feedback is a key step to ensure that all participants perceived a consistent tactile feedback. As the calibration cannot be done automatically, participants were asked to calibrate the intensity of the electrotactile interpenetration feedback multiple times throughout the experiment.
If the calibration is not done per each user, parameters such as the skin's conductivity and/or a tolerance could result in stimuli that are not perceived (i.e., too low intensity) or uncomfortable or even painful (i.e., too high intensity).%

The right range of current intensity was determined using the method of limits. Starting from a low intensity value, the intensity was increased considering steps of 0.1~mA. This incremental procedure continued, step by step, until the participant confirmed that the electrotactile feedback was perceived, defining his/her sensation detection threshold. Then, the incremental approach continued until the participant confirmed that the electrotactile feedback caused discomfort, defining his/her pain/discomfort threshold. Similarly, starting from a high intensity value, the intensity was decreased considering steps of 0.1~mA. This starting high intensity value was defined as the 3rd quartile of the range between the detection and pain/discomfort thresholds.
This decremental procedure continued, step by step, until the participant confirmed that the electrotactile feedback was not perceived anymore.%

During the entire calibration process, the signal frequency was set to 200~Hz and the pulse width to 500~$\mu$s. These values were chosen as they represent the strongest stimulation that could be achieved through the interpenetration feedback, as detailed in the previous Section. The final intensity value to be used throughout the experiment was calculated from a linear interpolation between the detection and discomfort thresholds. We chose an intermediate value between these thresholds, as it would result in a perceivable yet comfortable stimulus%

\subsection{Participants and Apparatus}
A total of  21 individuals [age: $mean(SD) = 25.67(2.74)$; $range = 23 - 33$ years old; gender: 14 males, 6 females, and 1 non-binary] were recruited through an in-campus advertisement and a call on institutional email lists. 21 participants were familiar with using VR head-mounted displays (VR-HMDs),
10 participants had a previous experience with receiving tactile feedback in VR (predominantly vibrotactile), and only 2 had experienced electrotactile feedback in the past. 

An HTC Vive Pro HMD was used for immersion. The PC had an NVIDIA RTX 2070. An HTC Vive Tracker attached to the back of the palm (see \autoref{fig:system}) and 2 Lighthouse base stations were used for tracking.
See~\autoref{fig:system} for the placement of electrotactile stimulator, which weighted 300~g. A Velcro strap was used to hold the electrode, its tightness was adjusted to achieve a comfortable yet firm placement of the electrode.

\begin{figure}[t!]
 \centering 
 \includegraphics[width=0.9\columnwidth,trim=5cm 0 5cm 2cm,clip]{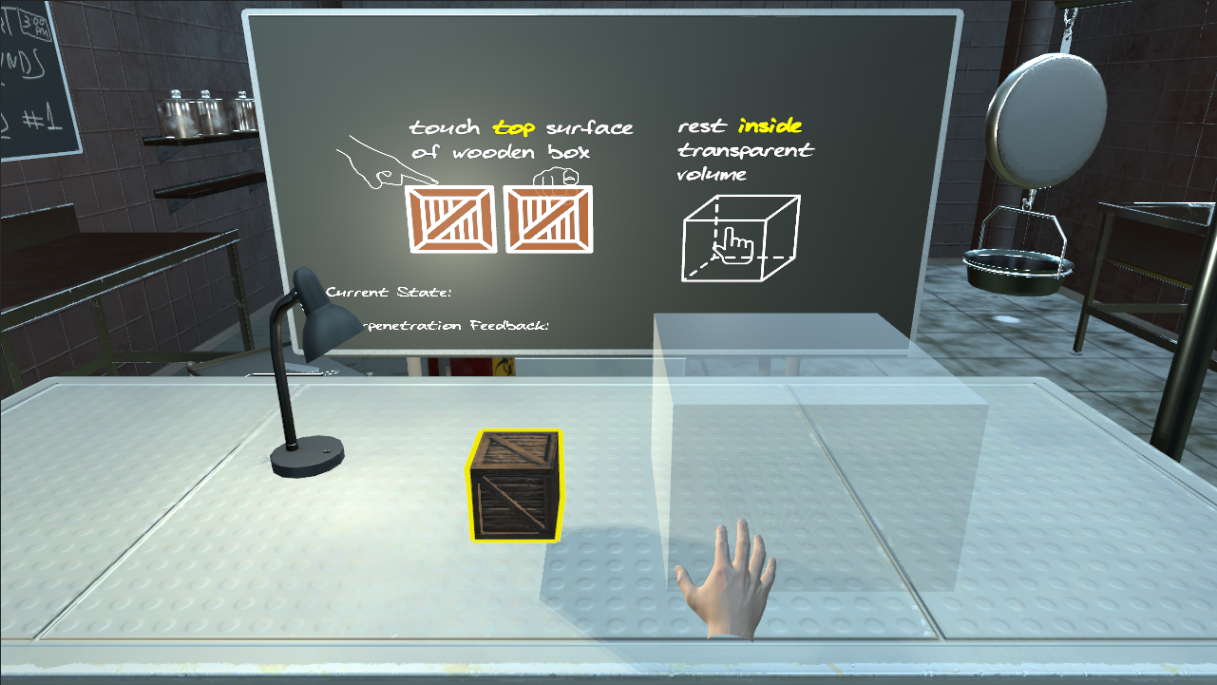}
 \caption{Depiction of the subjective view of the user during the experiment. The user could touch the virtual cube placed on the top of the table of the left. The semitransparent cube on the right side defines the resting area. The blackboard in front of the user provided general information during the experiment.}
 \label{fig:environment}
\end{figure}

The VR application was built using Unity in conjunction with the SteamVR Unity plugin (see~\autoref{fig:environment}).
The collisions were directly handled by the SteamVR plugin and the Unity physics simulation.
A blackboard was used for providing any critical information/instructions within the participant's field of view. 
\subsection{User Representation and Visual Interpenetration feedback}
\label{sec:visFeedback}
\begin{figure}[t!]
  \includegraphics[width=.45\columnwidth]
    {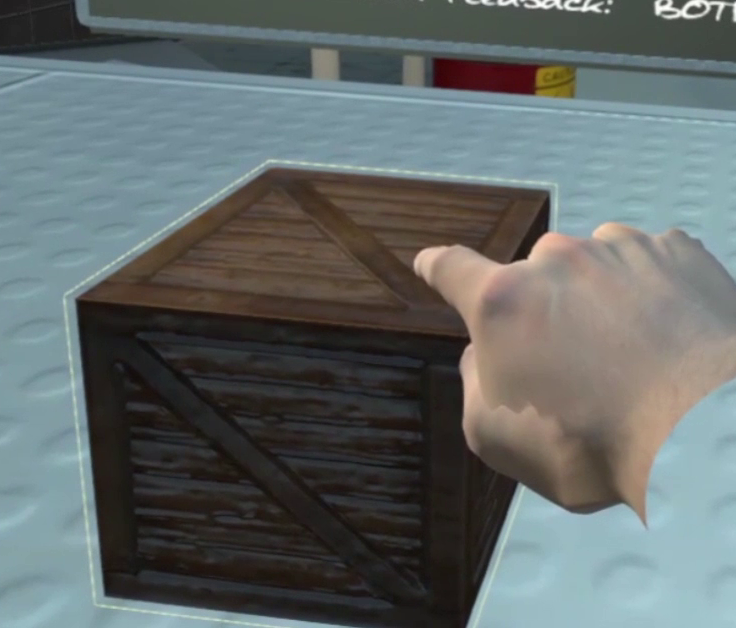}\hfill
  \includegraphics[width=.45\columnwidth]
    {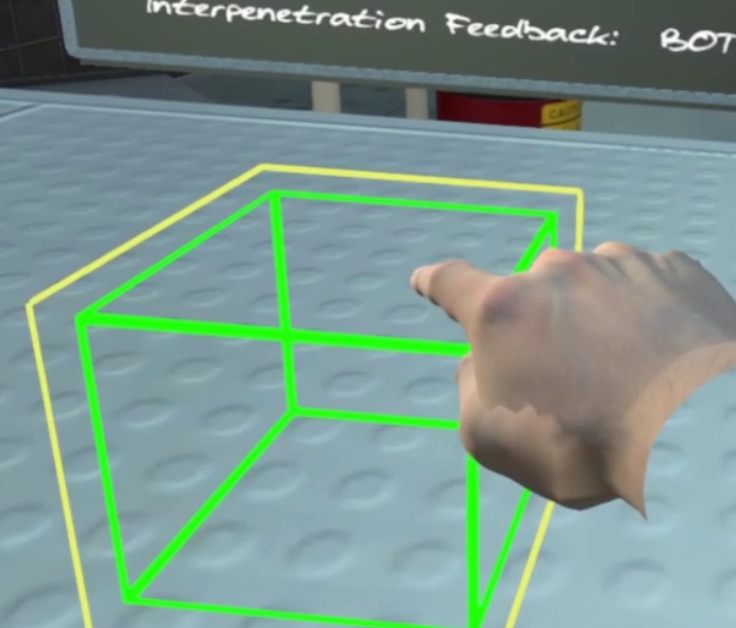}
  \caption{Visual shading conditions (left \& right) and visual interpenetration feedback (right).}
  \label{fig:shadingVFeedback}
\end{figure}

The user could only see a visual representation of their right hand (proxy/avatar). As stated earlier, the hand avatar cannot go through the virtual object so as to better mimic a real interaction. The hand is always kept in a closed fist, with the index finger fully extended, which is the same pose we asked participants to keep while interacting with the cube (see \autoref{fig:shadingVFeedback}).%

An outline effect was used as visual interpenetration feedback (see~\autoref{fig:shadingVFeedback}. Outline effects are commonly used to denote interactivity or selection \cite{Zhao2019, Sidenmark2020}. However, the implemented outline was dynamically rendered proportional to interpenetration's depth. The outline was parameterized using a scale factor and the size of the border. The scale factor went from $\times1$ (edge of $15$~cm$^3$) to $\times1.2$ (edge of $18$~cm$^3$). The border 
varied between 1 and 5 pixels.

When both electrotactile and visual interpenetration feedbacks were provided, a mismatch between the two modalities may result in a haptics' uncanny valley effect, which compromises the performance~\cite{berger2018uncanny}. For this reason, the modification of border size and width of the outline was determined through a pilot study. The choice of this particular parameters' range aimed to ensure that, comparably to the electrotactile interpenetration feedback, the perception of the visual interpenetration feedback's intensity is proportional to the size of the interpenetration.%

\subsection{Experimental Protocol}
This study received the approval of the authors' institution ethical committee.
As it was carried out during the COVID-19 pandemic, a preventive hygiene and disinfection protocol was implemented. 
Participants signed an informed consent form prior to their participation
The set up consisted of a VR area, a table, and a chair. 
The table was co-located with the virtual table.
A pre-exposure questionnaire was filled to acquire demographic data.
Participants went through a tutorial for performing the experimental task. 

The experiment followed a latin square design, \textit{4 types of interpenetration feedback} $\times$ \textit{2 types of cube shading} $\times$ \textit{2 parts of the experiment} (i.e., $4\times2\times2$). 
Each experiment's part had 4 blocks, where each block consisted of 12 repetitions.
In total, each participant performed 96 repetitions (i.e., 48 per experiment's part).
In total, 3 calibrations were performed a) Before Part 1; b) After Part 1; c) After Part 2. 
Finally, a post-exposure questionnaire was filled to acquire the participant's subjective ratings on the types of feedback, with a focus on electrotactile interpenetration feedback.

\subsection{Dependent Variables and Hypotheses}
In order to answer the hypothesis of our experiment and considering the experimental task we measured the following metrics:
average ($avg\_d$), standard deviation ($std\_d$) and maximum ($max\_d$) interpenetration during the three seconds in which the user was asked to remain in contact with the virtual cube surface.
As we did not track the user's finger but the back of the hand, the interpenetration was computed by measuring the distance between the position of the \emph{real} fingertip in the virtual environment and the position of the avatar fingertip constrained on the surface of the cube.
The interpenetration measures provide information regarding the awareness of the real interpenetration (accuracy) and their ability to keep a constant and not oscillatory contact with the virtual surface (precision).
In this respect our hypotheses were:
\begin{itemize}
    \item \textbf{[H1]} The addition of interpenetration feedback (either visual or electrotactile) will increase the precision and the accuracy of the task.
    \item \textbf{[H2]} The combination of the visual and electrotactile interpenetration feedback will result on the best precision and accuracy.
    \item \textbf{[H3]} Precision and accuracy will be higher for the fully shaded condition.
    \item \textbf{[H4]} Precision and accuracy will improve along the experiment due to learning effects.
\end{itemize}
The sensation and discomfort thresholds (in mA) were obtained during the three calibration processes. We expected that the thresholds will vary due to changes in skin conductivity or due to habituation.
In this respect our hypothesis was:
\begin{itemize}
    \item \textbf{[H5]} Sensation and discomfort thresholds will monotonically increase during the experiment.
\end{itemize}
Finally, the participants' subjective report were collected through a post-exposure questionnaire. The questionnaire is included as supplementary material. The questionnaire encompassed questions pertaining to the perceived usefulness, coherence, and resemblance with a real interaction of the electrotactile interpenetration feedback, as well as some comparison questions between the interpenetration feedback modalities.
In this respect our hypotheses were:
\begin{itemize}
    \item \textbf{[H6]} Electrotactile interpenetration feedback will be reported as useful and coherent.
    \item \textbf{[H7]} There will be no significant differences between visual and electrotactile interpenetration feedback in terms of usefulness, coherence, and/or resemblance.
\end{itemize}
\section{Results}
The homogeneity and normality assumptions were examined. The performance variables violated the normality assumption. The values were converted into logarithms since the variables' distributions were substantially positively skewed. No violation was then detected. The same amendments were made for the intensity values, which also violated the normality assumption. 
Finally, as the shading condition showed no significant results, which suggests the rejection of \textbf{H3}. In sake of clarity, the shading condition will not be considered in the results report. Thus, for the following analyses the conditions were the type of interpenetration feedback and the part of experiment.

A two-way repeated measures ANOVA was performed for each performance variable, as well as a post-hoc pairwise comparisons test respectively. When the sphericity assumption was violated, then the Greenhouse Geisser's correction was considered. 
Furthermore, a pairwise comparison was conducted to explore potential differences in the intensity values amongst the three calibrations (i.e., initial, middle, and final). 
Also, a Pearson's correlational analysis was performed to investigate the presence of relationships between the intensity values and the performance variables. 
Finally, the post-exposure self-reports (Likert scale) by the participants were explored. A non-parametric pairwise comparison was implemented to examine the presence of differences on the questionnaire's items. The Bonferroni correction was implemented accordingly to the number of comparison's pairs in each analysis to amend p-values' inflation due to multiple comparisons. 
The descriptive statistics for all performance variables and the intensities, as well as the results for the maximum interpenetration are all included in the supplementary material for completeness.

\subsection{Index Finger Interpenetration}
\textbf{Average Index Interpenetration - Accuracy}
The ANOVA's outcomes indicated that all effects were statistically significant. The type of feedback produced a main effect of $F(1.97,39.31) = 25.89$, $p <.001$ postulating a substantial difference amongst the feedback types. In line with \textbf{H1}, an $\mathcal{\omega}^2 =  0.54, 95\%$ $CI[0.35,0.66]$ suggesting a very large effect of the type of feedback on average interpenetration (i.e., accuracy). In support of \textbf{H4}, the part of the experiment yielded an effect of $F(1,20) = 31.74, p <.001$, suggesting a significant difference between the two parts of the experiment with a very large effect size of $\mathcal{\omega}^2 = 0.58, 95\%$ $CI[0.26,0.75]$. The interaction between the parts of the experiment and the types of feedback produced an effect of $F(2.11,42.21) = 5.72$, $p =.006$ indicating significant differences among the pairs between the types of feedback and the parts of experiment. An $\mathcal{\omega}^2 = 0.18, 95\%$ $CI[-0.01,0.34]$ was yielded for these interactions indicating that they have a large effect on the accuracy. 

\begin{figure}[!t]
 \centering 
 \includegraphics[width=0.95\columnwidth]{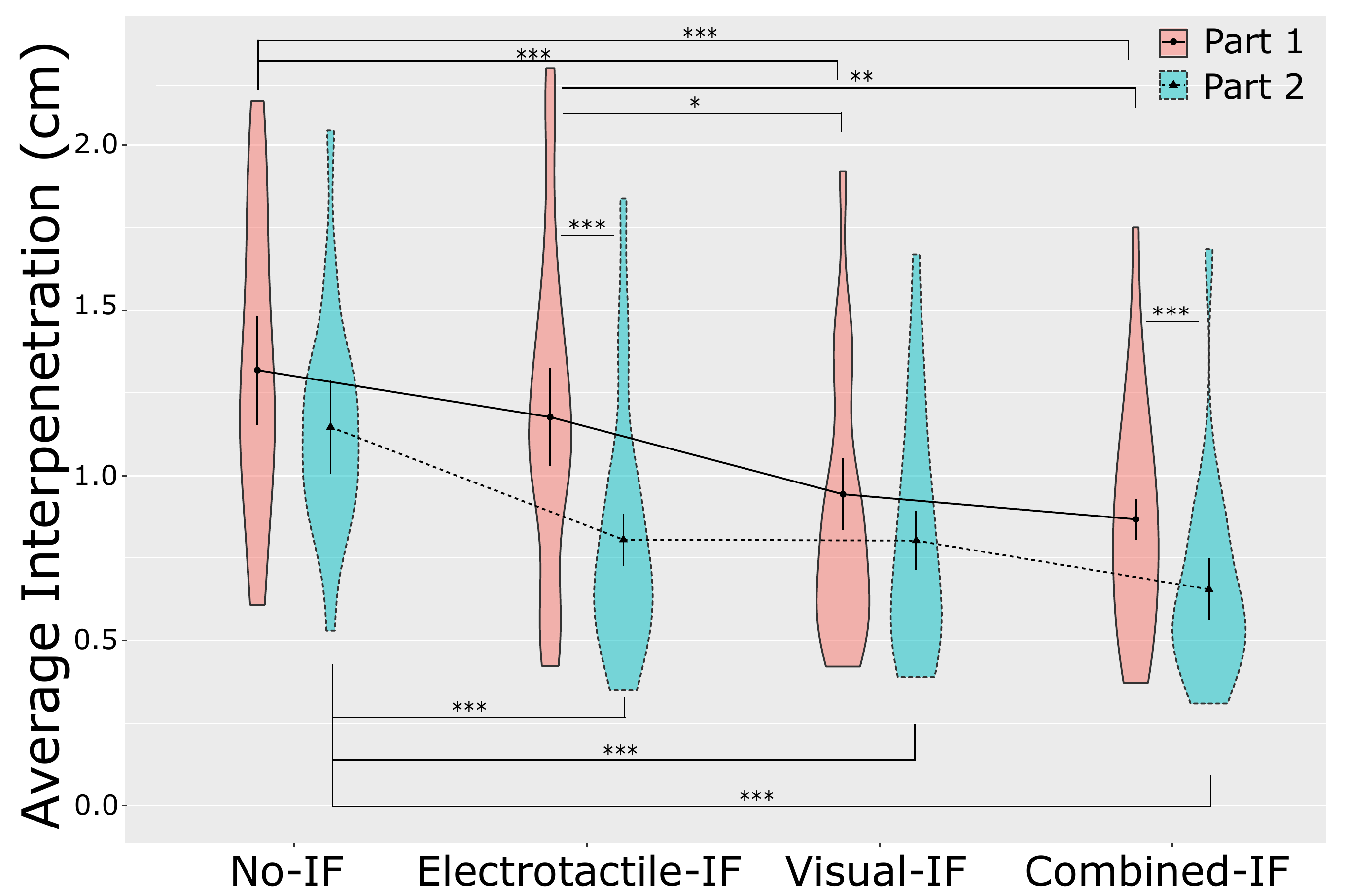}
 \caption{Average Interpenetration in cm. Comparisons per Type of Feedback and Part of the Experiment. IF = Interpenetration Feedback}
 \label{fig:AvInt}
\end{figure}

The post-hoc pairwise comparisons are displayed in \autoref{fig:AvInt}. The results support both \textbf{H1} and \textbf{H2}. In the 1$^{st}$ part, the visual $g = 0.90, p <.001$ and combined feedback $g = 1.10, p <.001$ were substantially improved compared to the no interpenetration feedback condition, with a large effect in both comparisons. Also, the visual $g = 0.52, p =.026$ and combined feedback $g = 0.67, p =.002$ were substantially better than the electrotactile condition, with a moderate effect size in both comparisons. In the 2$^{nd}$ part, the visual $g = 1.10, p <.001$, combined $g = 1.69, p <.001$, and electrotactile interpenetration feedback $g = 1.10, p <.001$, were significantly better compared to the no interpenetration feedback condition, with a very large effect in every comparison. No further significant differences were detected in both parts. Furthermore, the electrotactile $g = 0.86, p <.001$ and combined feedback $g = 0.69, p <.001$ were significantly improved in the 2$^{nd}$ part compared to the 1$^{st}$ part, with a large and a moderate effect correspondingly. In contrast, the equivalent comparisons for the no interpenetration feedback and visual interpenetration feedback were not statistically significant. Notably, only the size of change for electrotactile interpenetration feedback was significantly greater than the size of change for the no interpenetration feedback $g = 0.64, p =.005$ and visual interpenetration feedback $g = 0.60, p =.010$, with a moderate effect size for both.

\textbf{Standard Deviation Index Interpenetration - Precision}
The results of the ANOVA suggested statistically significant effects at .05 significance level. Supporting \textbf{H1}, the main effect of the the type of feedback on index oscillations (i.e., precision) yielded an effect of $F(1.88,37.69) = 28.92, p <.001$, and of $\mathcal{\omega}^2 = 0.57, 95\%$ $CI[ 0.39,0.68]$ postulating significant differences between the diverse types of feedback with a very large effect size. Alligned with \textbf{H4}, the main effect for the part of experiment on the precision produced an $F(1,20) = 9.32, p =.006$, and of $\mathcal{\omega}^2 =  0.27, 95\%$ $CI[-0.01, 0.54]$ suggesting a significant difference between the two part of the experiment with a large effect. The interaction among the diverse pairs derived from the parts of the experiment and the types of feedback yielded an interaction effect of $F(2.38,47.66) = 4.30, p =.014$, and of $\mathcal{\omega}^2 = 0.13, 95\%$ $CI[-0.03,0.29]$ postulating significant difference between the diverse pairs with a large effect size.

\begin{figure}[t!]
 \centering 
 \includegraphics[width=0.95\columnwidth]{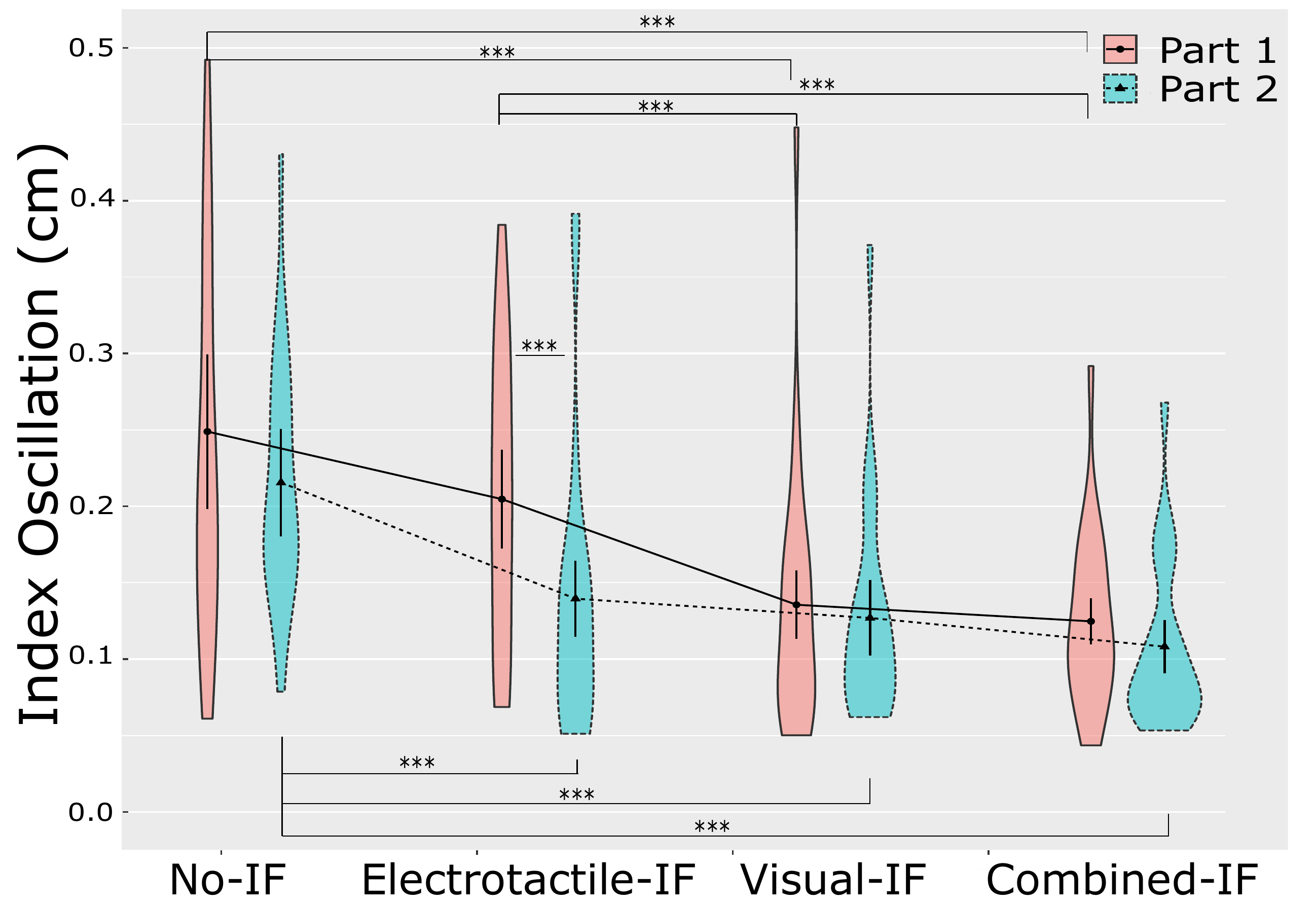}
 \caption{Index Oscillation in cm. Comparisons per Type of Feedback and Part of the Experiment. IF = Interpenetration Feedback}
 \label{fig:IndOsc}
\end{figure}

The post-hoc paiwise comparisons are illustrated in \autoref{fig:IndOsc}. The results appear to support both \textbf{H1} and \textbf{H2}. In the 1$^{st}$ part of experiment, the visual $g = 1.06, p <.001$ and combined feedback $g = 1.109, p <.001$ were founded substantially greater than the no interpenetration feedback condition, with a very large effect in both comparisons. Also, the visual $g = 0.79, p <.001$ and combined feedback $g = 0.90, p <.001$ were found to be substantially better than the electrotactile interpenetration feedback, with a large effect size in both comparisons. In the 2$^{nd}$ part of experiment, the visual $g = 1.32, p <.001$, combined $g = 1.71, p <.001$, and electrotactile interpenetration feedback $g = 1.07, p <.001$, were found to be significantly better than the no interpenetration feedback condition, with a very large effect in every comparison. No further significant differences were detected in both parts. Furthermore, only the electrotactile $g = 0.74, p <.001$ was found to be substantially improved in the 2$^{nd}$ part compared to the 1$^{st}$ part, with a large effect. The equivalent comparisons for the no interpenetration feedback, visual interpenetration, and combined interpenetration feedback were not statistically significant. Importantly, the size of change for electrotactile interpenetration feedback from the 1$^{st}$ to the 2$^{nd}$ part was significantly larger than the size of change of the no interpenetration feedback $g = 0.54, p =.027$ and visual interpenetration feedback $g = 0.59, p =.012$, with a moderate effect size in both comparisons.

\subsection{Calibration of Electrotactile Feedback} 
\begin{figure}[t!]
 \centering 
 \includegraphics[width=0.95\columnwidth]{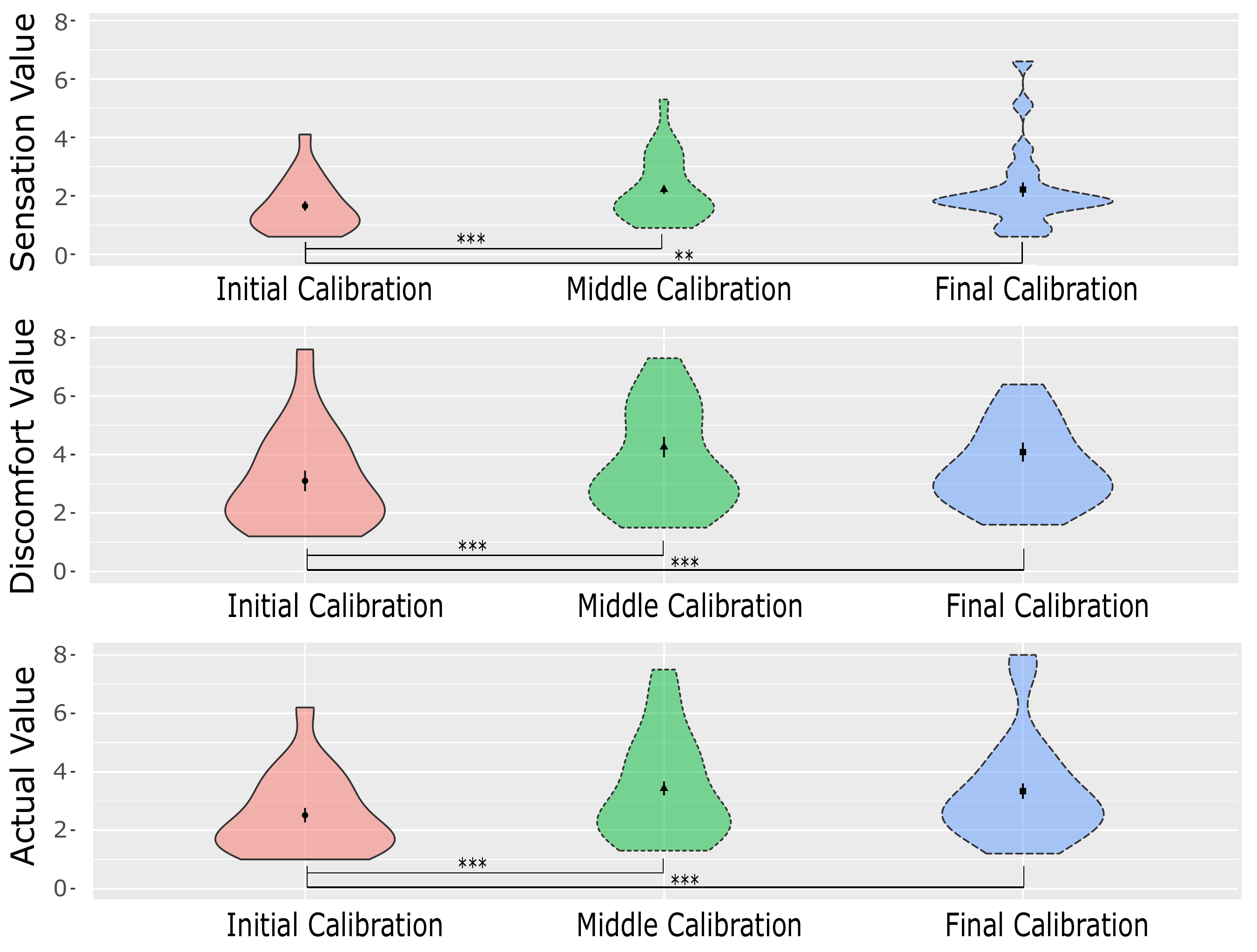}
 \caption{Intensity values in mA. Comparisons for each calibration.}
 \label{fig:Intensities}
\end{figure}

The pairwise comparisons are displayed in \autoref{fig:Intensities}. The sensation threshold (i.e., the intensity that the participant commences perceiving the electrotactile feedback), the discomfort threshold (i.e.,intensity that the participant commences feeling discomfort due to electrotactile feedback), and the actual value (i.e., the weighted intensity that was implemented to provide electrotactile interpenetration feedback) of the middle (i.e., after the 1$^{st}$  and before the 2$^{nd}$ part of experiment) and final calibration (i.e., after the 2$^{nd}$ part of experiment) were statistically greater than the equivalent intensities of initial calibration (i.e., before the 1$^{st}$ part of experiment). Specifically, the sensation threshold $g = 1.45, p <.001$, the discomfort threshold $g = 0.91, p <.001$, and the actual value $g = 1.08, p <.001$ of middle calibration were substantially larger than the equivalent intensities of initial calibration, while the effect size was large to very large in each comparison. Similarly, the sensation threshold $g = 0.70, p =.005$, the discomfort threshold $g = 0.83, p <.001$, and the actual value $g = 0.88, p <.001$ of final calibration were significantly larger than the equivalent intensities of initial calibration, while the effect size was large in every comparison. However, no significant differences were detected between final and middle calibration, which postulates the rejection of \textbf{H5}.  Finally, none of the intensity values was significantly correlated either positively or negatively with any of the performance variables.%

\subsection{Subjective Results}%
The descriptive statistics of the post-exposure self-reports are displayed in \autoref{tab:Questionnaire}, which support \textbf{H6}. In support of \textbf{H7}, the Wilcoxon signed rank test yielded no significant differences in the comparisons between the visual and electrotactile interpenetration feedback pertaining to usefulness (i.e., how useful was the feeback for the task's completion), coherence (i.e., how much coherent was the feedback with touching the surface), and resemblance (i.e., how much the feedback resembles touching a real surface). The Wilcoxon rank sum test, for the question pertaining to the combined feedback condition, indicated a significant difference $p = .024$ between electrotactile and combined feedback, postulating that the participants reported that they relied significantly more on the electrotactile interpenetration feedback rather than both interpenetration feedbacks (i.e., visual and electrotactile). However, in the same question, there was no significant difference between visual and electrotactile interpenetration feedback suggesting a balanced preference.%

\begin{table}[h]
    \caption{Self-reports on Electrotactile Interpenetration Feedback}
    \addtolength{\tabcolsep}{1pt}
        \centering
        \begin{tabular}{cccc}
        \small
        {\textbf{Criterion}} & \small{\textbf{Median}} & \small{\textbf{Mode}} & \small{\textbf{Range}}\\ \hline
        \small{Usefulness}        & \small{6}      & \small{5} & \small{2 - 7} \\
        \small{Coherence}         & \small{5}     & \small{5} & \small{4 - 7}  \\
        \small{Resemblance}       & \small{2.5}    & \small{1}    & \small{1 - 6}  \\
        \small{Pleasantness}      & \small{4}     & \small{4}    & \small{3 - 6}  \\
        \small{Perceptual Update} & \small{5.5}    & \small{6}    & \small{2 - 7}  \\
        \small{Combined Synchronicity}     & {6}      & {7}    & \small{2 - 7}  \\ \hline
        \multicolumn{4}{c}{\small{\textit{N = 21; Max = 7; $\ge4$ indicate positive response;}}}\\
        \multicolumn{4}{c}{\small{\textit{Median = middle number of the provided responses;}}}\\
        \multicolumn{4}{c}{\small{\textit{Mode = most frequent response}}}
        \end{tabular}
    \label{tab:Questionnaire}
\end{table}
Moreover, the participants reported that the visual interpenetration feedback was useful and coherent, albeit that provided a feeling that is different from touching a real surface. Similarly, it was reported that the electrotactile interpenetration feedback was useful to very useful and coherent, albeit that provided a feeling that is different to very different from touching a real surface. Also, the self-reports suggested that the electrotactile interpenetration feedback provided neither a pleasant nor unpleasant sensation, as well as that it provided a frequent to very frequent perceptual update of pressing down the surface. Finally, the self-reports indicated that the combination of electrotactile and visual interpenetration feedback had a very coherent to completely coherent synchronization, and it was perceived as moderately similar as to touching a real surface.%
\section{Discussion}
Our findings agree with previous studies, where bimodal feedback increased the accuracy and precision in contact tasks \cite{Triantafyllidis2020}. Also, our results are in accordance with a meta-analysis of user interaction studies, where the bimodal (i.e., visuohaptic) feedback was found to improve performance \cite{burke2006comparing}. However, a haptic uncanny effect may substantially compromise the benefits of bimodal feedback on performance~\cite{berger2018uncanny}. In our study, the dynamic visual interpenetration feedback was matched to the varying intensities of electrotactile interpenetration feedback, which aligns with the suggestions to mitigate or evade a haptic uncanny effect \cite{berger2018uncanny}. Hence, our results further support the importance of having matched feedback modalities, when implementing bimodal feedback.%

Furthermore, our quantitative and qualitative outcomes align with the qualitative results of Pamungkas, et al. \cite{Pamungkas2013}, where the participants reported that the electrotactile interpenetration feedback assisted them with performing a VR robot-teleoperation task faster and more accurately. However, in our quantitative user study the benefits of providing electrotactile interpenetration feedback became apparent only in the 2$^{nd}$ part of experiment, suggesting the presence of a learning effect. Also, the intensity values for providing electrotactile interpenetration feedback were substantially increased in the middle calibration, which were used in the 2$^{nd}$ part of experiment. In conjunction, our findings postulate that the advantages of electrotactile interpenetration feedback may surface, when the user has been familiarized with it, and/or the intensity of feedback is well-adjusted to skin's conductivity changes and/or user's increased tolerance to the electrotactile interpenetration feedback. However, the intensity values were not correlated with neither accuracy nor precision. Thus, the increased familiarity with the electrotactile interpenetration feedback that was inferred due to the observed learning effect, appears to better explain why the advantageous effects of electrotactile interpenetration feedback on performance were surfaced only during the 2$^{nd}$ part of experiment.%

It should be noted that the learning effect was not generally present. The learning effect was identified only for the bimodal visuotactile feedback regarding accuracy, as well as for the electrotactile interpenetration feedback regarding both accuracy and precision. These findings indicate that the learning effect was important only in conditions where the electrotactile interpenetration feedback was used. Also, the size of the improvement in the accuracy and precision from the 1$^{st}$ to the 2$^{nd}$ part of experiment was significantly greater for the electrotactile interpenetration feedback condition compared to the size of improvement for no interpenetration feedback and visual interpenetration feedback conditions respectively. This finding also postulates that the learning effect was observed due to the novelty of the electrotactile interpenetration feedback, and not due a practice effect related to the task. Nevertheless, the effects of the familiarity with electrotactile interpenetration feedback and the refined calibration on user's performance should be further scrutinized in prospective studies.%

Finally, the results postulate the effectiveness of electrotactile and/or visual interpenetration feedback. However, the implemented visual interpenetration feedback requires constant visual focus, would occlude other visual information, and render visual artefacts in a fully interactive virtual environment, which may reduce substantially the contact precision and accuracy \cite{qian2015virtual}. In contrast, the suggested electrotactile interpenetration feedback appears to facilitate multiple contact interactions, although this should be further examined. Thus, the presented electrotactile feedback system seems to be the most appropriate and effective choice for improving precision and accuracy of mid-air interactions in immersive VR%

\section{Conclusion}%
In this paper, we proposed an electrotactile rendering algorithm to enhannce contact information in a virtual environment based on the modulation of the pulse width and frequency of the electrical signal. In order to explore the benefits and limitations of the proposed method, we conducted a user study evaluating the electrotactile interpenetration feedback by examining the performance in touching accurately and precisely a virtual rigid object.%

The results showed that participants achieved significantly better performance in terms of accuracy and precision when visual and/or electrotactile interpenetration feedback was provided. More importantly, there were no significant differences between visual and electrotactile interpenetration feedback when the calibration was optimized and the user was familiarized with electrotactile interpenetration feedback. Considering that visual discrimination is dominant in such scenarios and a visual interpenetration feedback renders visual artefacts, the findings indicate a promising future for electrotactile-based systems for improving contact precision and accuracy in interactive VR systems.%

Further works could explore other alternatives for the modulation and location of the electrotactile feedback or consider other tactile actuators. Also, future work should attempt to replicate the experimental findings with finger tracking capabilities. Further experiments could explore interactions requiring more fingers (e.g. grasp operation) or the interaction with dynamic objects (e.g. pushing a virtual object). Finally, further iterations of the system should strive to improve its wearability.%

\section*{Acknowledgments}
This work was supported by the European Union’s Horizon 2020 research and innovation program under grant agreement No. 856718 (TACTILITY).

\bibliographystyle{eg-alpha-doi.bst}
\bibliography{my}

\end{document}